\documentclass{cupconf}

  \checkfont{eurm10}
  \iffontfound
    \IfFileExists{upmath.sty}
      {\typeout{^^JFound AMS Euler Roman fonts on the system,
                   using the 'upmath' package.^^J}%
       \usepackage{upmath}}
      {\typeout{^^JFound AMS Euler Roman fonts on the system, but you
                   dont seem to have the}%
       \typeout{'upmath' package installed. cupconf.cls can take advantage
                 of these fonts,^^Jif you use 'upmath' package.^^J}%
      }
  \else
  \fi

  \checkfont{msam10}
  \iffontfound
    \IfFileExists{amssymb.sty}
      {\typeout{^^JFound AMS Symbol fonts on the system, using the
                'amssymb' package.^^J}%
       \usepackage{amssymb}%

      }{}
  \fi

  \IfFileExists{amsbsy.sty}
    {\typeout{^^JFound the 'amsbsy' package on the system, using it.^^J}%
     \usepackage{amsbsy}}
    {\providecommand\boldsymbol[1]{\mbox{\boldmath $##1$}}}

\newsavebox{\astrutbox}
\sbox{\astrutbox}{\rule[-5pt]{0pt}{20pt}}

\title[The Star Formation History in the Andromeda Halo]{The Star Formation
History in the Andromeda Halo}

\author[T.M. Brown]%
{T\ls H\ls O\ls M\ls A\ls S\ns M.\ns B\ls R\ls O\ls W\ls N}

\affiliation{Space Telescope Science Institute, 3700 San Martin Drive, 
Baltimore, MD 21218}

\pubyear{2003}

\begin{document}

\maketitle

\begin{abstract}
I present the preliminary results of a program to measure the star
formation history in the halo of the Andromeda galaxy.  Using the
Advanced Camera for Surveys (ACS) on the Hubble Space Telescope, we
obtained the deepest optical images of the sky to date, in a field on
the southeast minor axis of Andromeda, 51$^\prime$ (11 kpc) from the
nucleus.  The resulting color-magnitude diagram (CMD) contains
approximately 300,000 stars and extends more than 1.5 mag below the
main sequence turnoff, with 50\% completeness at $V=30.7$~mag.  We
interpret this CMD using comparisons to ACS observations of five
Galactic globular clusters through the same filters, and through
$\chi^2$-fitting to a finely-spaced grid of calibrated stellar
population models.  We find evidence for a major ($\sim 30$\%)
intermediate-age (6--8 Gyr) metal-rich ([Fe/H]$>-0.5$) population in
the Andromeda halo, along with a significant old metal-poor population
akin to that in the Milky Way halo.  The large spread in ages
suggests that the Andromeda halo formed as a result of a more violent
merging history than that in our own Milky Way.

\end{abstract}

\firstsection 
\section{Introduction}

One of the primary quests of observational astronomy is understanding
the formation history of galaxies.  An impediment to this research is
the relative paucity of galaxies in the Local Group, which contains no
giant ellipticals, and only two giant spirals -- our own Milky Way and
Andromeda.  Fortunately, Andromeda (M31, NGC224) is well-situated for
studying the formation of giant spiral halos, due to its proximity
(770 kpc; \cite{FM90}), small foreground reddening
($E_{B-V}=0.08$~mag; \cite{SFD98}), and low inclination ($i \approx
12.5^{\rm o}$; \cite{dV58}).  Andromeda is similar to the Milky Way in
many respects (Hubble type, absolute magnitude, mass, and size;
\cite{vdB92}; \cite{KZS02}), but we have long known that its halo is
more metal-rich than that of the Milky Way; the metallicity
distribution in the Milky Way halo peaks near [Fe/H]$\approx -1.8$
(\cite{RN91}), while that in the Andromeda halo peaks near
[Fe/H]$\approx -0.5$ (\cite{MK86}; \cite{HFR96}; \cite{DHP01}).
Although the Milky Way halo is dominated by old stars (e.g.,
\cite{VdB00}), the formation history of the Andromeda halo has been
unknown until now.

Physical processes possibly at work in forming spiral halos include
rapid dissipative collapse in the early universe (\cite{ELS62}),
slower accretion of separate subclumps (\cite{L69}; \cite{SZ78}), 
and dissolution of globular clusters (\cite{AHO88}).  More recent
hierarchical models suggest that spheroids (bulges and halos) form in
a repetitive process during the mergers of galaxies and protogalaxies,
while disks form by slow accretion of gas between merging events
(e.g., \cite{WF91}).  The discovery of the Sgr dwarf galaxy
(\cite{IGI94}) sparked renewed interest in halo formation
through accretion of dwarf galaxies, and ambitious programs are now
underway to map out the spatial distribution, kinematics, and chemical
abundance in the halos of the Milky Way (e.g., \cite{MMOH00};
\cite{MOK00}) and Andromeda (e.g., \cite{FII02}).
In the meantime, the realization that hierarchical models based on
cold dark matter almost inevitably predict many more dwarf galaxies
than are actually seen around the Milky Way (\cite{MGG99}) has
led to suggestions that most of the dwarf galaxies formed in the early
Universe have dissolved into the halo (e.g., \cite{BKW00}).
Whether or not such accretion is the dominant source of stars in the
stellar halo, it is likely that dwarf galaxies do contribute, and at
large galactocentric distances their stars can remain in coherent
orbital streams for many Gyr.  Indeed, one such stream has been found
in the halo of Andromeda (\cite{IIL01}; \cite{MII03}); 
\cite[Ferguson et al.\ (2002)]{FII02} demonstrated dramatic evidence for
large-scale debris trails and substructure in the M31 halo and outer disk,
yet their deep imaging and star counts also found an apparently smooth 
component extending to large radii.

The most direct way to measure ages in a stellar population is to
construct a color-magnitude diagram (CMD) that reaches to stars below
the main sequence turnoff.  For decades, researchers have used such
data to determine the ages of Galactic globular clusters and satellite
galaxies of the Milky Way, but until now there has been no instrument
capable of resolving these stars in a massive galaxy outside of our
own.  With the installation of the Advanced Camera for Surveys (ACS;
\cite{FBB98}) on the Hubble Space Telescope (HST), this is now
possible.  We have obtained deep ACS observations of a minor axis
field in Andromeda's halo, $\approx 51$ arcmin (11 kpc) from the
nucleus.  In these proceedings, I present the preliminary results of
our program (\cite{B03}).  I start with a brief review of the
age diagnostics available in a deep CMD, then describe our ACS
observations and their analysis, and finish with the implications for
the formation history of Andromeda.

\section{Age Diagnostics}

\begin{figure}
\includegraphics{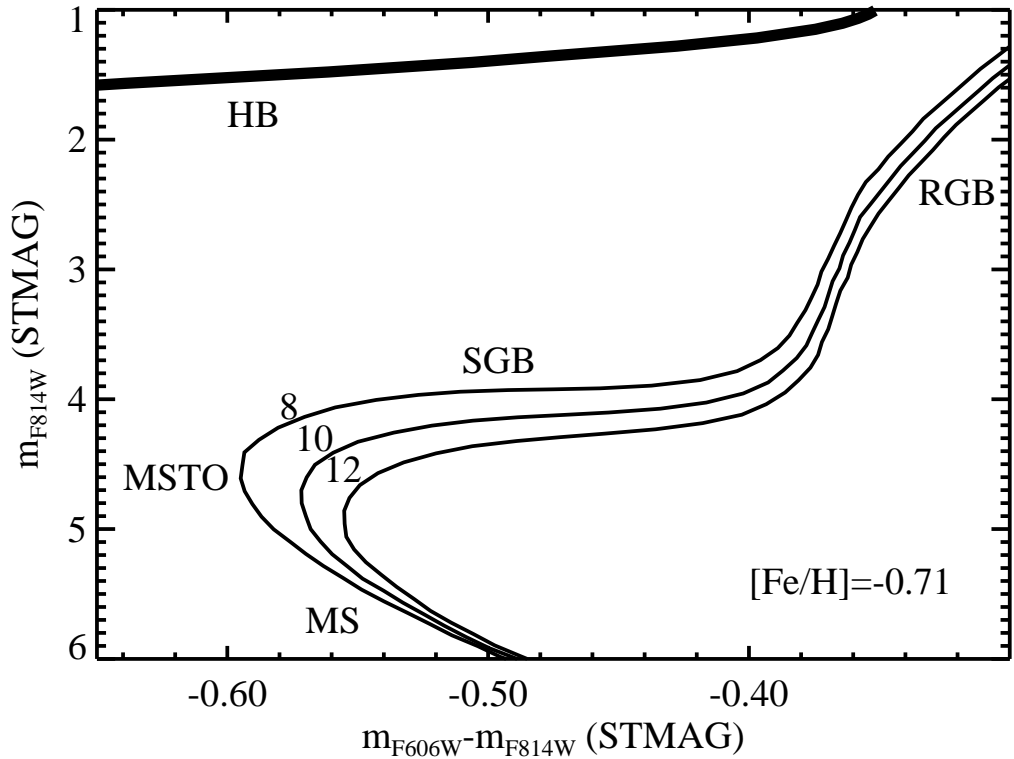}
\vspace{4.1in}
\caption{The variation in optical photometry of a stellar population
with age at fixed metallicity ({\it labeled}).  Isochrones at 8, 10, and 12
Gyr ({\it labeled}) are shown in a CMD constructed using the ACS bandpasses
F606W and F814W.  Note that the subgiant branch (SGB) and main sequence
turnoff (MSTO) are more sensitive to age than the red giant branch (RGB).
The primary age indicators in the CMD are the luminosity difference
between the MSTO and the horizontal branch (HB), which becomes larger
as age increases, and the color difference between the MSTO and the RGB,
which becomes smaller as age increases.  However, in order to clearly
define the MSTO, photometry must extend to the main sequence (MS) stars
below the MSTO.}
\label{agediag}
\end{figure}

Stellar population ages are best determined from resolved optical
photometry reaching the main sequence.  Although the location of the
main sequence turnoff in the CMD is the primary age indicator,
photometry reaching well below the turnoff is required
to define the turnoff accurately.  Ages among the
Galactic globular clusters are usually determined via one of two
diagnostics (figure~\ref{agediag}): the luminosity difference between the 
turnoff and the horizontal branch (\cite{S82}; \cite{IR84}),
and the color difference between the turnoff and the red
giant branch (\cite{VBS90}).  The luminosity of the horizontal branch
and the color of the red giant branch are both relatively insensitive
to age (at fixed composition), while the main sequence turnoff becomes
both fainter and redder at increasing age.  Our ACS observations used
the F606W (broad $V$) and F814W ($I$) filters; for roughly every Gyr
beyond 10~Gyr, the main sequence turnoff becomes fainter by $\sim
0.1$~mag in $m_{F814W}$ and redder by $\sim 0.01$~mag in
$m_{F606W}-m_{F814W}$ (see figure~\ref{agediag}).

\begin{figure}
\includegraphics{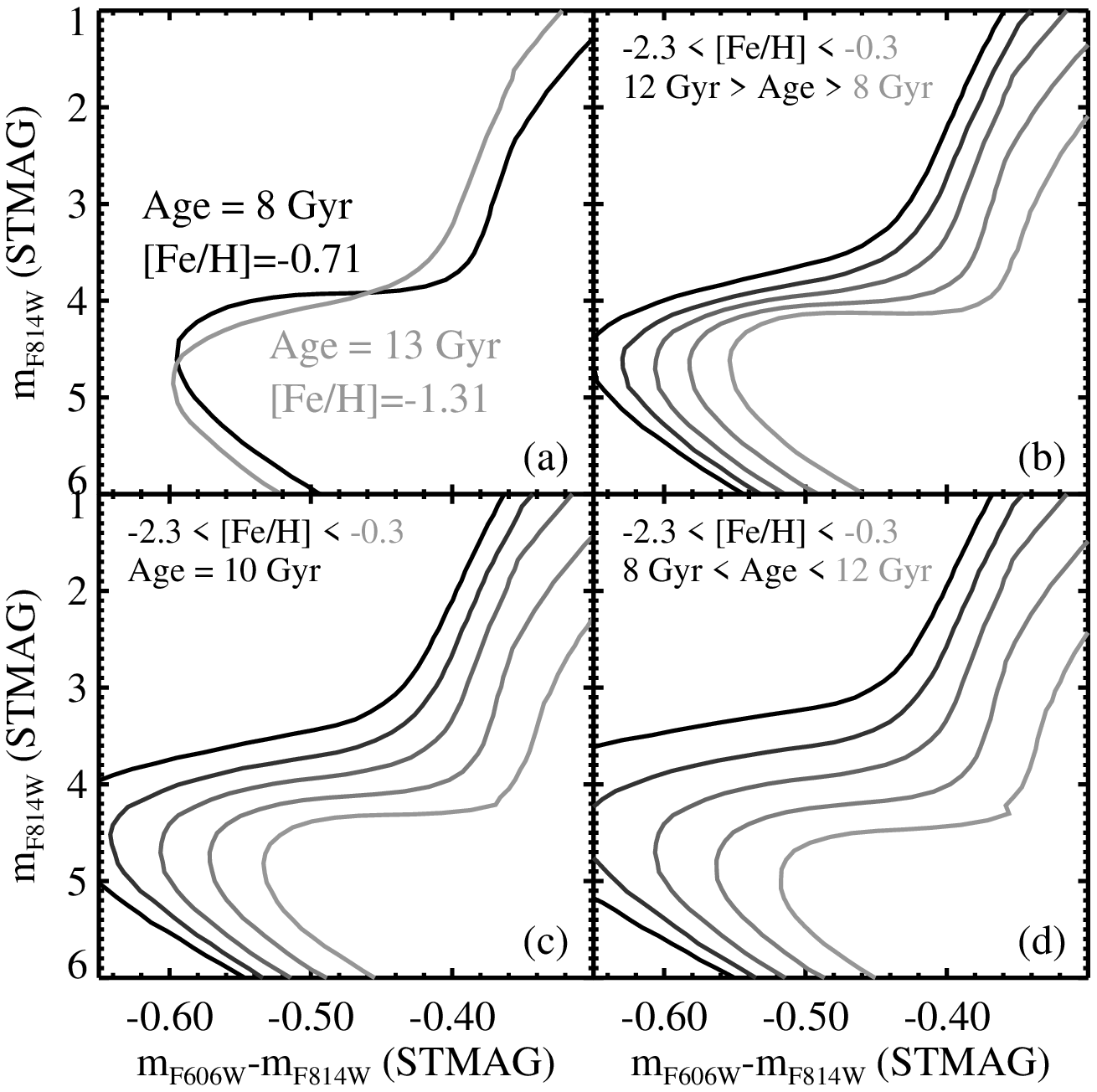}
\vspace{5.4in}
\caption{{\it Panel a:} An isochrone of intermediate age and
metallicity ({\it dark curve, labeled}) and an older, more metal-poor
isochrone ({\it light curve, labeled}) intersect, demonstrating that
two-color photometry for a single star does not uniquely determine age
and metallicity.  Both isochrones also have the same
color at the main sequence turnoff, and thus unresolved populations
with these parameters would appear very similar.  However, resolved
photometry could distinguish between two such populations. {\it Panel
b:} Isochrones spanning 8-12 Gyr at 1 Gyr intervals, with a range in
metallicity ({\it labeled}) anticorrelated with age, such that the youngest
isochrone is the most metal-rich, as might be expected for a simple
model of chemical evolution.  The subgiant branch and main sequence
turnoff both become fainter and redder at increasing age or increasing
metallicity, while the red giant branch is more sensitive to
metallicity than age.  Because the effects of age and metallicity
counteract each other at the subgiant branch, it appears very narrow
compared to the red giant branch.  {\it Panel c:} The same
metallicities as in {\it b}, but all of the isochrones have the same
age.  {\it Panel d:} The same metallicities and ages as in {\it b},
but now age is correlated with metallicity, such that the older
populations are more metal-rich.  This is somewhat unphysical, but
demonstrates a situation of renewed star formation from the infall of
primordial gas.  Note the width of the subgiant branch compared to the
red giant branch.  Panels {\it b--d} demonstrate that resolved
photometry of a mixed population can disentangle age and metallicity,
because each evolutionary phase responds differently to these parameters.
}
\label{agemetdeg}
\end{figure}

The age-metallicity degeneracy is a major uncertainty when
characterizing individual stars via photometry, or unresolved
populations via low-resolution spectra (figure~\ref{agemetdeg}a).
However, the situation is better for a resolved stellar
population.  Because the red giant branch is much more sensitive to
metallicity than age, and the subgiant branch and main sequence
turnoff are sensitive to both age and metallicity, the distribution of
stars on the red giant branch, subgiant branch, and main sequence
turnoff can be simultaneously reproduced only for specific
combinations of age and metallicity (figure~\ref{agemetdeg}).

\section{Observations}

\begin{figure}
\includegraphics{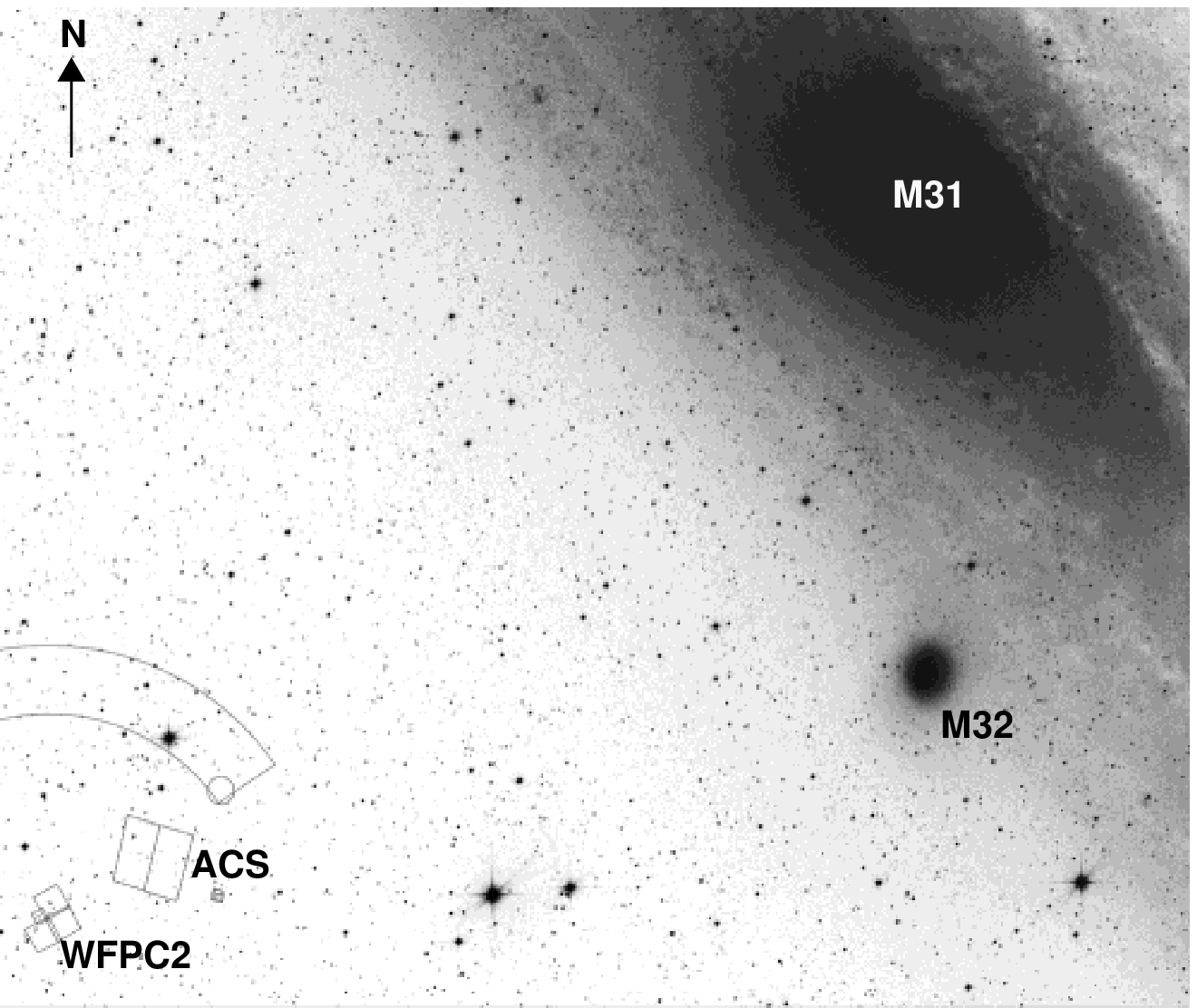}
\vspace{4.6in}
\caption{A $50^\prime \times 60^\prime$ Digital Sky Survey image of
Andromeda, showing the positions of our ACS and WFPC2 images
(labeled). The ACS field is 50 arcmin (11 kpc) from the center of M31.
The nearly edge-on ($i \approx 12.5^{\rm o}$; \cite{dV58}) disk should
contribute $\lesssim 3$\% of the stars in the ACS field.}
\label{fieldpos}
\end{figure}

Using the ACS Wide Field Camera (WFC), we obtained deep optical images
of a field along the southeast minor axis of the M31 halo, at
$\alpha_{2000} = 00^h46^m07^s$, $\delta_{2000} = 40^{\rm
o}42^{\prime}34^{\prime\prime}$ (figure~\ref{fieldpos}).  The field
was previously imaged by \cite[Holland et al.\ (1996)]{HFR96} with the
Wide Field Planetary Camera 2 (WFPC2); the field is not associated
with the tidal streams and substructure found by \cite[Ferguson et
al.\ (2002)]{FII02}, and lies just outside the ``flattened inner
halo'' in their maps.  Given the nearly edge-on disk, the contribution
of disk stars at this position should be $\lesssim 3$\% (\cite{WK88};
\cite[Holland et al.\ 1996]{HFR96} and references therein). We chose
this field to optimize the crowding (trading off population statistics
versus photometric accuracy) and to place an interesting M31 globular
cluster (GC312; \cite{SKH77}) near the edge of our images.  The
metallicity of GC312 ([Fe/H]$=-0.7$; \cite{HBK91}) is near the peak in
the metallicity distribution for the M31 halo; this should simplify
our attempts to derive relative ages for this cluster and the halo
(currently underway).  Due to scheduling and orientation constraints,
one bright foreground star ($V\sim 14$~mag) was unavoidable; the star
and its window reflection affect a few percent of the total image
area, which we discard.

\begin{figure}
\includegraphics{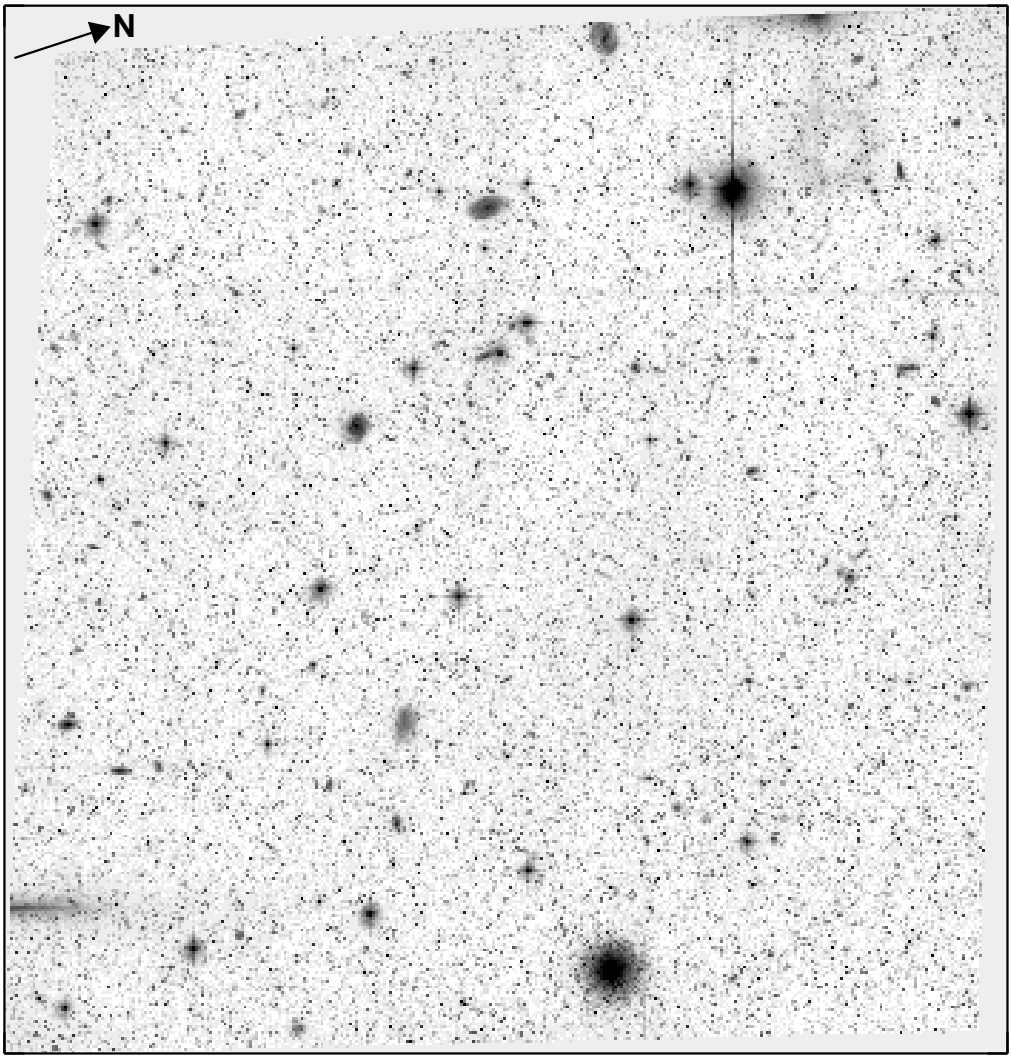}
\vspace{5.6in}
\caption{The ACS F606W image of the Andromeda halo, shown at a log
stretch, and binned for display purposes by a factor of 20, to
0.6$^{\prime\prime}$ pixel$^{-1}$. The field subtends $211^{\prime\prime}
\times 221^{\prime\prime}$ (not much larger than the WFC field of
view, because the largest dither was about 6$^{\prime\prime}$).  For
our preliminary analysis, we excluded the area around GC312 ({\it bottom})
and around a bright ($V=14$~mag) foreground star ({\it upper right}) and its
window reflection (to the right of the star).  Many background
galaxies can be seen through the 300,000 stars detected in Andromeda's
halo, but they are a small ($< 5$\%) contamination.  The image has
been corrected for the strong geometric distortion in the camera,
giving the field its unusual shape.}
\label{acshalo}
\end{figure}

From 2 Dec 2002 to 11 Jan 2003, we obtained 39.1 hours of ACS images
in the F606W filter (broad $V$) and 45.4 hours in the F814W filter
($I$), with every exposure dithered to allow for hot pixel removal,
optimal sampling of the point spread function, smoothing of the
spatial variations in the detector response, and filling the gap
between the two halves of the $4096 \times 4096$ pix detector.  We
co-added the M31 images using the IRAF DRIZZLE package, with masks for
the cosmic rays and hot pixels, resulting in geometrically-correct
images with a plate scale of 0.03$^{\prime\prime}$ pixel$^{-1}$ and an
area of approximately $210^{\prime\prime} \times 220^{\prime\prime}$
(figure~\ref{acshalo}). We then performed both aperture and
PSF-fitting photometry using the DAOPHOT-II package (\cite{S87}),
assuming a variable PSF constructed from the most isolated stars in
the images.  The aperture photometry on isolated stars was corrected to
true apparent magnitudes using TinyTim models of the HST PSF
(\cite{K95}) and observations of the standard star EGGR 102 (a
$V=12.8$~mag DA white dwarf) in the same filters, with agreement at
the 1\% level.  The PSF-fitting photometry was then compared to the
corrected aperture photometry, in order to derive the offset between
the PSF-fitting photometry and true apparent magnitudes. Our
photometry is in the STMAG system: $m= -2.5 \times $~log$_{10}
f_\lambda -21.1$.  For those more familiar with the Johnson $V$ and
Cousins $I$ bandpasses, a 5,000~K stellar spectrum has $V - m_{F606W}
= -0.05$~mag and $I - m_{F814W} = -1.28$~mag.

\begin{figure}
\includegraphics{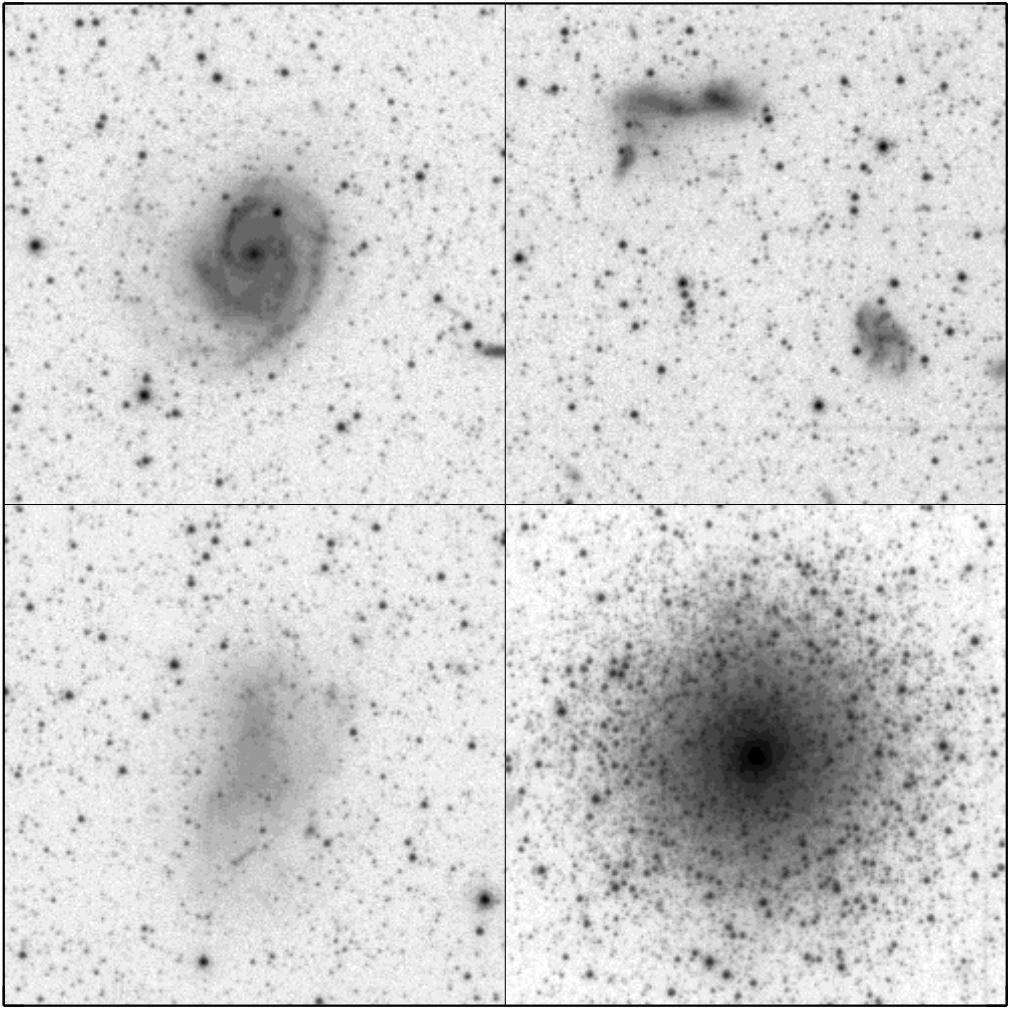}
\vspace{5.5in}
\caption{Subsections ($15^{\prime\prime} \times 15^{\prime\prime}$) of
the ACS image show the globular cluster GC312 in Andromeda's halo
({\it lower right}), background galaxies, and the level of crowding in
the halo population.  The images have been rebinned to a scale of
$0.06^{\prime\prime}$ pixel$^{-1}$ for display purposes.}
\label{haloinsets}
\end{figure}

Of the $\sim$300,000 stars detected, we discarded those within the
GC312 tidal radius (10$^{\prime\prime}$; \cite{HFR97}), within
14.5$^{\prime\prime}$ of a bright foreground star, within
12.6$^{\prime\prime}$ of this star's window reflection, and near the
image edges, leaving $\approx$223,000 stars in the final catalog.
Using the SExtractor code (\cite{BA96}), we estimate $\lesssim 5$\% of
the stars are contaminated by extended sources
(figure~\ref{haloinsets}).  Extensive
artificial star tests determine the photometric scatter and
completeness as a function of color and luminosity.  The CMD shows no
obvious differences when comparing the population in a
10--100$^{\prime\prime}$ annulus around GC312 to that beyond
100$^{\prime\prime}$; the cluster does not appear to be associated
with an extended underlying system.  By integrating our catalog, we
estimate that the surface brightness in our field is $\mu_V \approx
26.3$~mag arcsec$^{-2}$.

\begin{figure}
\includegraphics{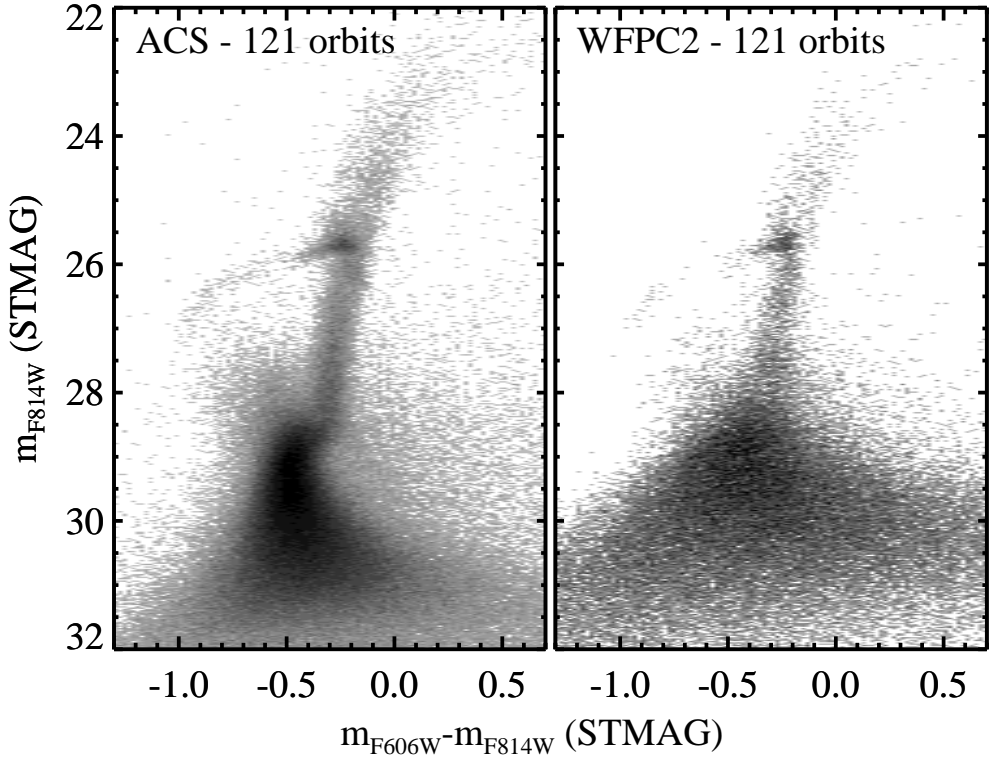}
\vspace{4.1in}
\caption{{\it Left panel:} The CMD constructed from the ACS images of
the Andromeda halo, using aperture photometry.  There are too many
stars to plot them individually; instead, this is a Hess diagram at a
logarithmic stretch.  {\it Right panel:} The CMD constructed from the
parallel WFPC2 images, in a field a few arcmin further out along the
minor axis of Andromeda, again using aperture photometry.  The filters
and exposure times in the WFPC2 data are approximately the same as
those in the ACS data.  Although the WFPC2 data reach stars at the
main sequence turnoff, they are not deep enough to characterize the
subgiant branch and the turnoff well.}
\label{figwfpc}
\end{figure}

We also obtained coordinated parallel WFPC2 observations of a second
field along the minor axis of Andromeda (figure~\ref{fieldpos}) using
the same bandpasses as those in the ACS observations.  The WFPC2 data
are not nearly as deep as the ACS data, but they are about as deep as
the Hubble Deep Field (\cite{WBD96}), and thus much deeper than any
Andromeda observations prior to our program.  Although the WFPC2
images resolve stars at the main sequence turnoff, the resulting CMD
is not deep enough to characterize the turnoff and subgiant branch
well.  The striking differences between the WFPC2 and ACS CMDs are a
testament to the technical advances achieved by the HST servicing
missions (figure~\ref{figwfpc}).  

\begin{figure}
\includegraphics{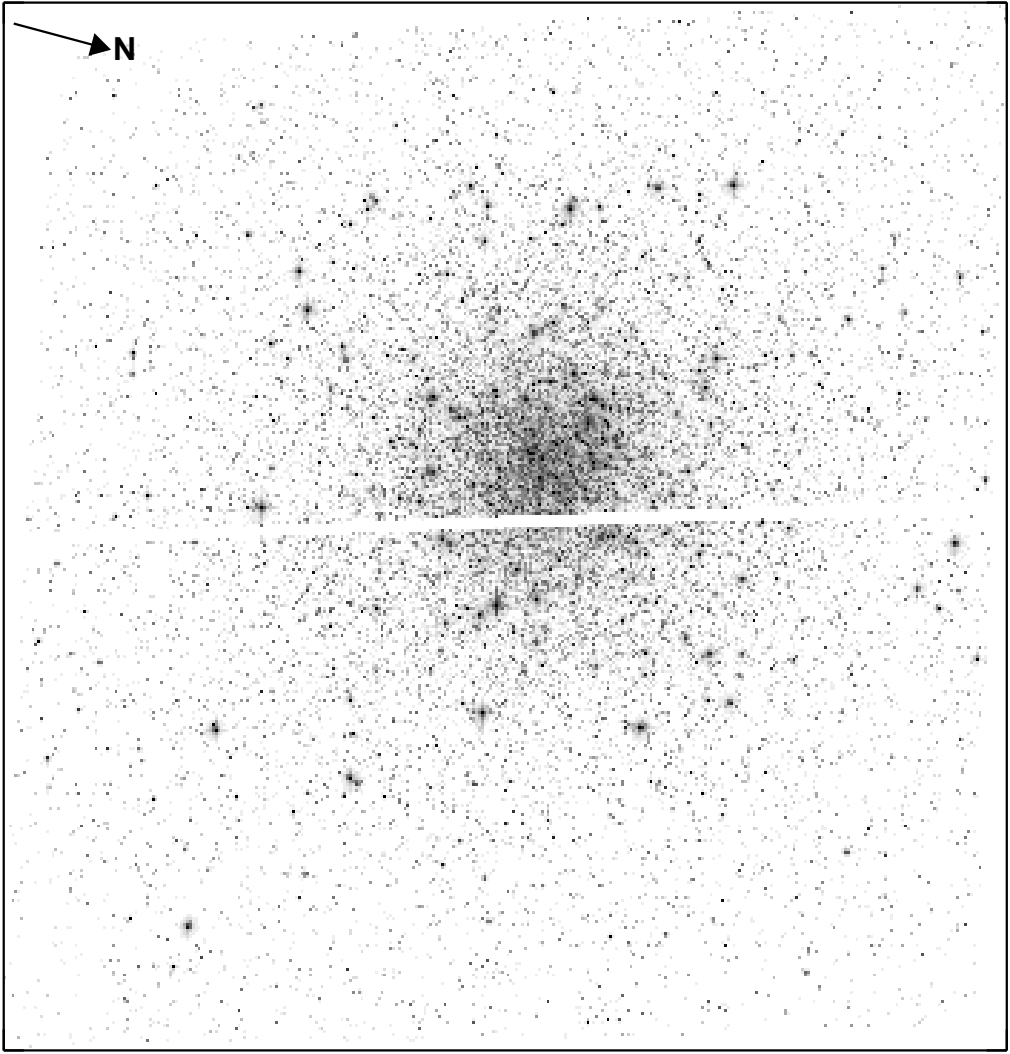}
\vspace{5.6in}
\caption{The ACS F606W image of M92, shown at a log stretch and binned
by 20 to $0.6^{\prime\prime}$ pixel$^{-1}$ for display purposes.  The
globular cluster images in our program were not dithered, so a gap
appears between the two WFC chips.}
\label{acsm92}
\end{figure}

\begin{figure}
\includegraphics{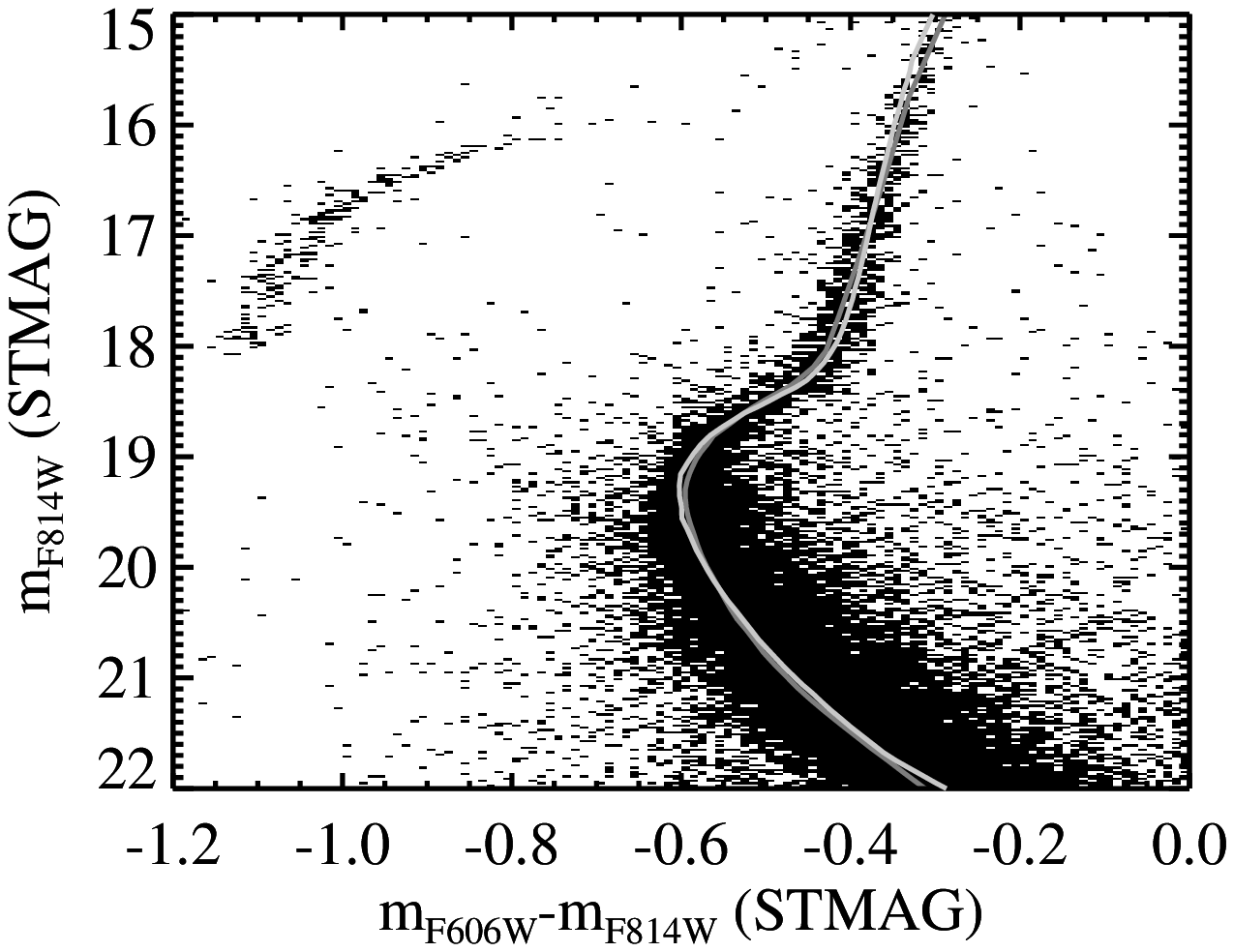}
\vspace{4.0in}
\caption{The CMD created from our ACS images of M92.  The ridge line
for this cluster ({\it dark grey curve}) was created by taking the
median $m_{F606W}-m_{F814W}$ color and median $m_{F814W}$ for a series
of points along the main sequence and red giant branch.  A 14~Gyr
isochrone at [Fe/H]=$-2.14$ ({\it light grey curve}), transformed to the
ACS bandpasses, agrees well with the ridge line over the CMD region
used for fitting the Andromeda halo population.}
\label{m92cmd}
\end{figure}

Because the ACS is a new instrument, our program includes ACS
observations of five Galactic globular clusters spanning a wide
metallicity range (Table 1), using the same filters utilized in our
Andromeda halo observations.  These cluster images allow the
construction of empirical isochrones in the ACS bandpasses, which can
be directly compared to the Andromeda CMD and used to calibrate the
transformation of the the theoretical isochrones to the ACS
bandpasses.  Figure~\ref{acsm92} shows the ACS F606W image of M92.
M92, NGC6752, and 47~Tuc are the most useful calibrators in our
program, because their parameters are known very well; NGC5927 and
NGC6528 are also useful because of their high metallicities, but their
parameters are less secure, and they suffer from high, spatially
variable foreground reddening (\cite{HR99}).  To increase the dynamic
range, three images were taken in each bandpass for each cluster, with
the exposure times varying by an order of magnitude.  To minimize the
number of orbits required for the cluster data, these observations
were not dithered.  Thus, we drizzled the data without plate scale
changes, in order to remove cosmic rays and to correct for geometric
distortion.  The cluster images are significantly less crowded (on
average) than those of M31, and the $0.05^{\prime\prime}$ pixels
undersample the PSF, so we performed aperture photometry but no PSF
fitting, and corrected the aperture photometry to true apparent
magnitudes (figure~\ref{m92cmd}).

\section{Analysis}

\begin{figure}
\includegraphics{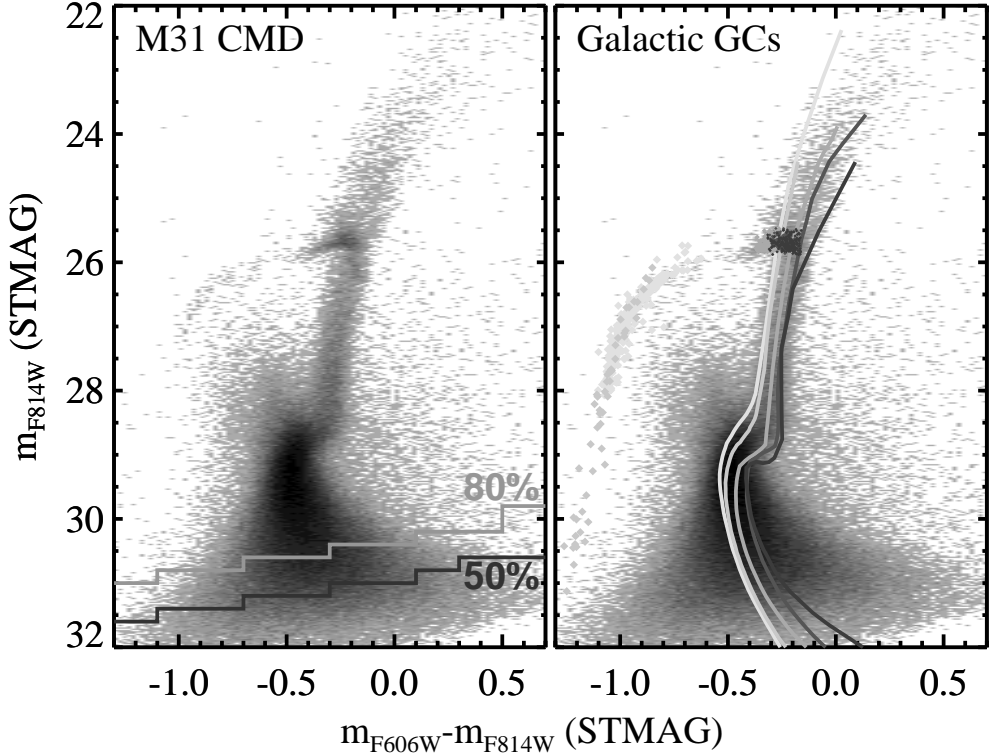}
\vspace{4.05in}
\caption{{\it Left panel:} The CMD of Andromeda constructed from the
PSF-fitting photometry of the ACS images.  Completeness limits
determined from artificial star tests are marked.  {\it Right panel:}
The CMD of Andromeda, with the ridge lines ({\it grey curves}) and
horizontal branch loci ({\it grey diamonds}) of five Galactic globular
clusters superimposed.  The shading of the ridge lines and horizontal
branch points range from light grey to dark grey as the cluster
metallicity increases (see Table 1).  The data for M92, NGC6752, and
47 Tuc have only been shifted by the differences in distance and
reddening between the clusters and M31, yet their horizontal branch
loci agree well with the horizontal branch of M31.  Those parameters
are very uncertain for NGC6528 and NGC5927, so the data for the two
metal-rich clusters were shifted to align their horizontal branch loci
with the M31 horizontal branch: NGC6528 was shifted 0.16~mag brighter,
while NGC5927 was shifted 0.11~mag brighter and 0.05~mag redder.
Moving from the most metal-poor cluster (M92) to the most metal-rich
(NGC6528), the red giant branches of the clusters span the width of
the red giant branch in M31, yet the subgiant branches of the clusters
become increasingly fainter than that of M31, indicating that the
metal-rich stars in the M31 halo are much younger than those in the
clusters.}
\label{comparegcs}
\end{figure}

The ACS CMD reveals, for the first time, the main sequence population in
the M31 halo. The horizontal branch extends from a well-populated red
clump to a minority blue population ($\sim 10$\% of the total
horizontal branch population).  The horizontal branch is not
noticeably extended -- the hot horizontal branch stars that are seen
in clusters like NGC6752 are missing.  The broad red giant branch
indicates a wide metallicity distribution extending to near-solar
metallicities, long-known to be characteristic of the M31 halo
(\cite{MK86}).  The luminosity function ``bump'' on the red giant
branch is another metallicity indicator, becoming fainter as
metallicity increases; it slopes away from the red horizontal branch
until the luminosity difference reaches $\approx$0.5~mag -- another
indication of near-solar metallicities.  There is also a prominent
blue plume of stars significantly brighter than the main sequence
turnoff; this minority population ($\sim 2$\% the size of the
population at the turnoff $\pm$1~mag) may include binaries, blue
stragglers, or a residual young stellar population.  Although the blue
horizontal branch stars are characteristic of very old, metal-poor
globular clusters (such as M92), the luminosity difference between the
turnoff and horizontal branch is smaller than expected for a purely
old stellar population.  This is shown in figure~\ref{comparegcs},
which shows the ridge lines and horizontal branch loci for the five
globular clusters we observed with ACS, superimposed upon the CMD of
Andromeda.  The M31 subgiant branch is nearly horizontal, indicating a
high metallicity, while its ridge is appreciably brighter than those
of 47~Tuc, NGC5927, and NGC6528, implying the presence of a
significantly younger population in the M31 halo.

The comparison between the globular clusters and Andromeda in
figure~\ref{comparegcs} provides a completely empirical indication of
the age spread in Andromeda's halo.  M92, NGC6752, and 47~Tuc
were shifted in color and magnitude according to the
differences in reddening and distance between the clusters and
Andromeda.  Those shifts naturally aligned the horizontal branch loci
of these clusters to the horizontal branch of Andromeda, thus
demonstrating the accuracy of the parameters in Table 1, but we could
have simply aligned the horizontal branches without any knowledge of
the relative distances and reddenings.  Indeed, forcing alignment at
the horizontal branch was required for NGC5927 and NGC6528,
because the parameters of those clusters are very uncertain.
Once the clusters are all aligned at the horizontal branch, the red
giant branches of the clusters span the broad red giant branch of
Andromeda, empirically demonstrating the wide metallicity distribution
in Andromeda's halo.  For the most metal-poor cluster (M92; {\it lightest
grey curve} in figure~\ref{comparegcs}), the
subgiant branch luminosity agrees well with that in Andromeda, and the
turnoff of the cluster agrees well with the blue edge of the turnoff
in Andromeda; this indicates that Andromeda contains a significant
population of metal-poor stars with ages similar to the oldest
Galactic globular clusters.  However, as one compares clusters at
increasing metallicity to Andromeda, discrepancies at the subgiant
branch and turnoff become more apparent.  Although the red giant branches
of 47~Tuc, NGC5927, and NGC6528 straddle a large fraction of the red
giant branch population in Andromeda, the cluster subgiant branches are well
below the bulk of the subgiant population in Andromeda.  This is a strong
indication that the metal-rich populations in Andromeda are significantly
younger than the cluster, given that the subgiant branch and turnoff become
fainter by approximately 0.1~mag for every 1 Gyr increase in age.

\begin{table}
\begin{center}
\caption{M31 and globular cluster parameters}
\begin{tabular}{cccccc}
\hline
        &           &           &  & \multicolumn{2}{c}{Exposure} \\
Name    & $(m-M)_V$ & $E_{B-V}$ &  & \multicolumn{2}{c}{(s)} \\
Name    & (mag)     &  (mag) & [Fe/H] & F606W & F814W \\
\hline
M31     & 24.68$^{\rm{a}}$ & 0.08$^{\rm{b}}$  & $-0.6$$^{\rm{c}}$ & 140870 & 163570  \\
\hline
M92     & 14.60$^{\rm{d}}$ & 0.023$^{\rm{b}}$ & $-2.14$$^{\rm{d}}$ & 95.5 & 106.5\\
NGC6752 & 13.17$^{\rm{e}}$ & 0.055$^{\rm{b}}$ & $-1.54$$^{\rm{f}}$ & 44.5 & 49.5\\
47~Tuc   & 13.27$^{\rm{g}}$ & 0.032$^{\rm{b}}$ & $-0.83$$^{\rm{f}}$ & 76.5 & 78 \\
NGC5927 & 15.81$^{\rm{h}}$ & 0.45$^{\rm{h}}$ & $-0.37$$^{\rm{h}}$ & 532 & 355.7\\
NGC6528 & 16.15$^{\rm{i}}$ & 0.55$^{\rm{i}}$  & $-0.2$$^{\rm{i}}$ & 504 & 371 \\
\hline
\end{tabular}
\end{center}
$^{\rm{a}}$\cite[Freedman \& Madore (1990)]{FM90}. 
$^{\rm{b}}$\cite[Schlegel et al.\ (1998)]{SFD98}. 
$^{\rm{c}}$\cite[Mould \& Kristian (1986)]{MK86}.
$^{\rm{d}}$\cite[VandenBerg \& Clem (2003)]{VC03}. 
$^{\rm{e}}$\cite[Renzini et al.\ (1996)]{R96}. 
$^{\rm{f}}$\cite[VandenBerg (2000)]{VdB00}.
$^{\rm{g}}$\cite[Zoccali et al.\ (2001)]{Z01}. 
$^{\rm{h}}$\cite[Harris (1996)]{H96}. 
$^{\rm{i}}$\cite[Momany et al.\ (2003)]{M03}.\\
\end{table}

\begin{figure}
\includegraphics{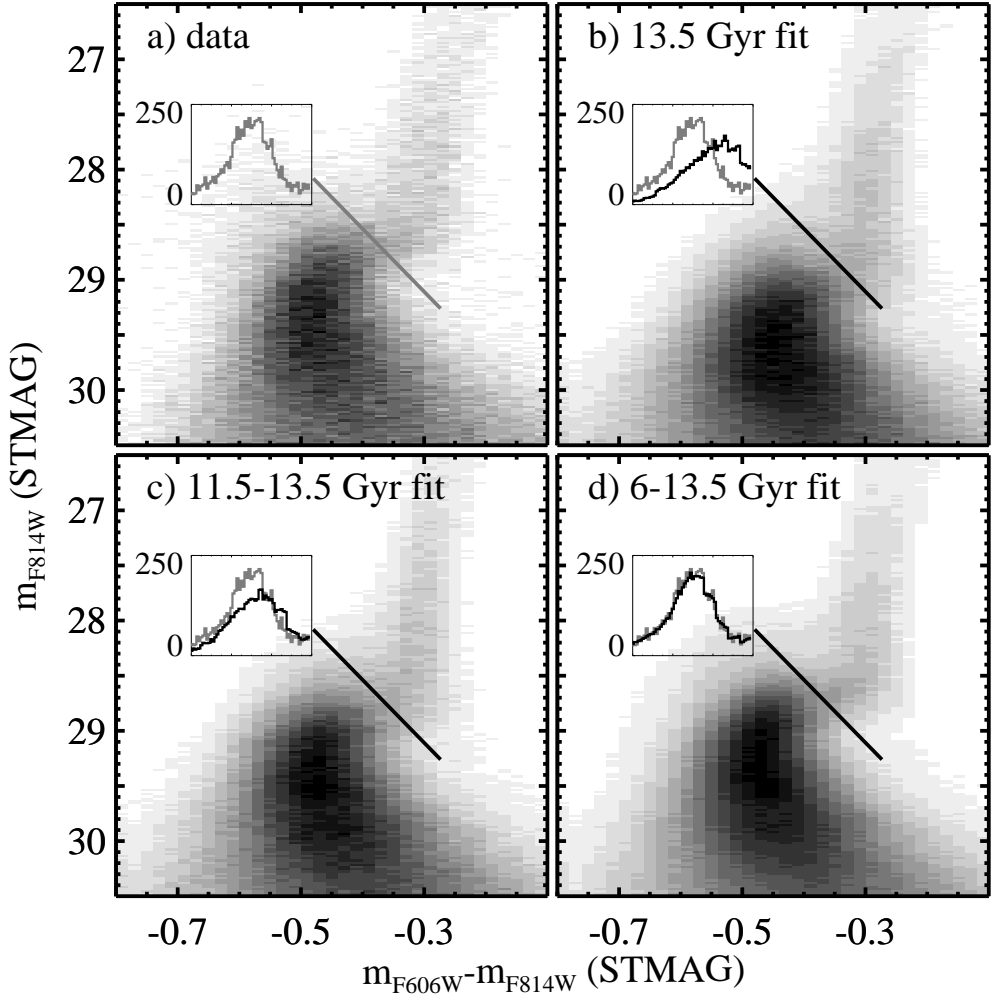}
\vspace{5.3in}
\caption{{\it Panel a:} The region of the Andromeda halo CMD that we used
for fitting the star formation history.  Restricting the fit to this
region avoids parts of the CMD that are seriously incomplete, sparsely
populated, or poorly constrained by the theoretical models.  A
histogram ({\it inset}) of the number of stars along a cut through the
subgiant branch ({\it grey line}) highlights the differences between
the data and the models in the subsequent panels.  {\it Panel b:} The fit to
the red giant branch using only 13.5 Gyr isochrones spanning a wide
range in metallicity.  Although such models can reproduce the width of
the Andromeda red giant branch, the subgiant branch and turnoff are
much fainter than those in the data.  The histogram ({\it inset})
compares the number of stars along the cut ({\it black line}) through
the subgiant branch to that in the data ({\it grey histogram}); they
are poorly matched.  {\it Panel c:} The fit to the entire region shown in
panel {\it a}, using isochrones with a wide metallicity range but
ages of 11.5--13.5 Gyr.  The histogram ({\it inset}) again highlights
that this model does not reproduce the distribution seen in the data.
{\it Panel d:} The fit to the data shown in panel {\it a}, using a wide
metallicity range and ages of 6--13.5 Gyr.  This model reproduces the
data well, as highlighted by the histograms ({\it inset}).  }
\label{fits}
\end{figure}

Our full analysis of the star formation history is in progress.
Here, I will supplement the comparisons shown in
figure~\ref{comparegcs} with examples from the modeling to date
(figure~\ref{fits}).  Our modeling is based upon the isochrones of
VandenBerg, Bergbusch, \& Dowler (2003, in
preparation), which show good agreement with cluster CMDs spanning a
wide range of metallicity and age (e.g., figure~\ref{m92cmd}).  To
transform these isochrones from the ground-based bandpasses to the
HST-based bandpasses, we used the spectra of \cite[Lejeune, Cuisinier,
\& Buser (1997)]{LCB97} to calculate $V - m_{F606W}$ and $I -
m_{F814W}$, as a function of effective temperature and gravity along
the isochrones, then applied those differences to the ground-based
magnitudes of the isochrones, with a small ($\lesssim 0.05$~mag)
empirical color correction to force agreement with our globular
cluster CMDs. After this correction, the isochrones match the ridge
lines within $\lesssim 0.02$~mag over the region of the CMD we are
fitting. Our observed CMDs of these clusters are reproduced by a
12.5~Gyr isochrone for 47~Tuc and 14~Gyr isochrones for NGC6752 and
M92.  The isochrones do not include core He diffusion, which would
reduce their ages by 10--12\%, thus avoiding discrepancies with the
age of the Universe (\cite{V02}).

Using isochrones with a range of ages and metallicities, we fit the
region of the M31 CMD shown in figure~\ref{fits}a using the StarFish
code of \cite[Harris \& Zaritsky (2001)]{HZ01}.  Restricting the fit
to this region of the CMD focuses on the most sensitive age and
metallicity indicators while avoiding regions of the CMD that are
seriously incomplete, sparsely populated, or poorly constrained by the
models (e.g., the horizontal branch morphology).  Note that we do not
include the red giant branch luminosity function bump in our fitting;
although theory predicts that this bump becomes fainter with
increasing metallicity, the theoretical zeropoint is fairly uncertain.
The StarFish code fits the observed CMD through a linear combination
of input isochrones, using $\chi^2$ minimization, where the isochrones
are scattered according to the results of the artificial star tests.
We first followed the standard method used when researchers determine
the metallicity distribution from the red giant branch
(figure~\ref{fits}b).  We fit the red giant branch using a set of old
(13.5~Gyr) isochrones spanning a wide metallicity range ($-2.31 < \rm
[Fe/H] < 0$).  The resulting metallicity distribution was similar to
that found by \cite[Holland et al.\ (1996)]{HFR96} in our same field,
and \cite[Durrell et al.\ (2001)]{DHP01} in a field 20 kpc from the
nucleus.  However, it is very obvious that these purely old isochrones
do not match the subgiant branch or the turnoff.  Next, as shown in
figure~\ref{fits}c, we tried fitting the entire region of
figure~\ref{fits}a with a wider age range (11.5--13.5 Gyr), but found
no acceptable combination.  The insets in figure~\ref{fits}
contain histograms showing the number of stars (data: {\it grey}; model:
{\it black}) along a cut through the subgiant branch ({\it thick
line}). In figure~\ref{fits}c, the residual subgiant stars that are not
reproduced by this old model ({\it inset}) suggest that at least 20\%
of the stars in the halo must be younger than 11.5~Gyr.  We conclude
that a purely old stellar population cannot explain the CMD of the M31
halo.

\begin{figure}
\includegraphics{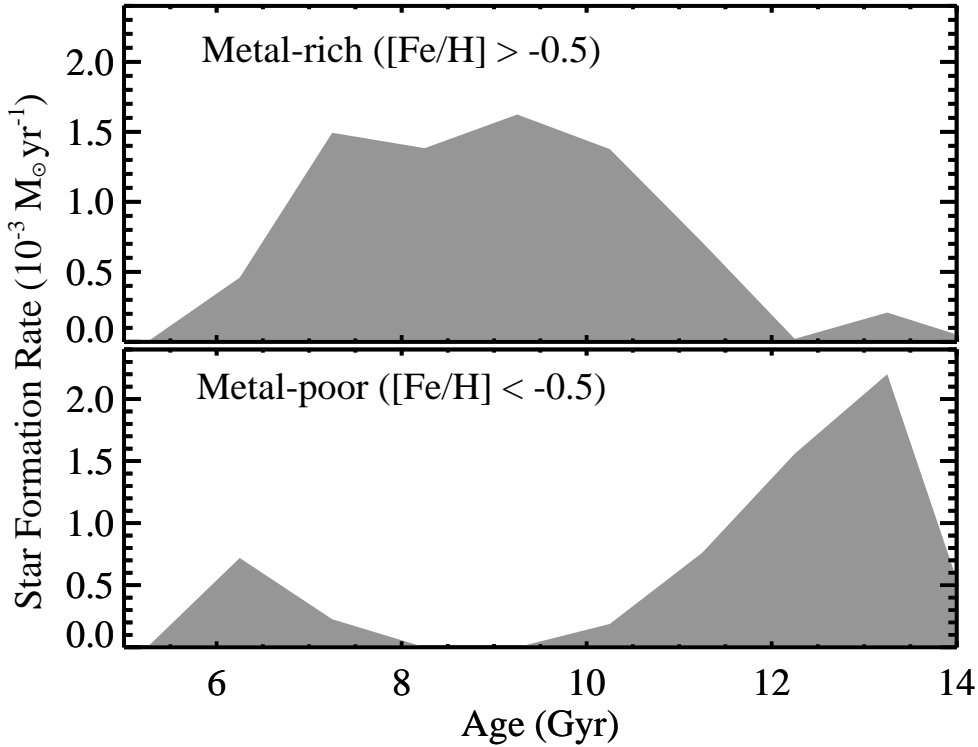}
\vspace{4.0in}
\caption{{\it Upper panel:} The star formation history of the metal-rich
stars in the best model (figure~\ref{fits}d) obtained to date.  The
bulk of the metal-rich population is of intermediate age.
{\it Lower panel:} The star formation history of the metal-poor stars
in the best-fit model.  Most of the metal-poor stars are old, but
this model includes a small population of young metal-poor stars.  Note
that this star formation history is not unique; we can obtain a fit
that is nearly as good as this fit, using two completely distinct
population components (a purely old metal-poor population and 
a purely young metal-rich population).  The detailed star formation
history will be explored more fully in a future paper.}
\label{sfh}
\end{figure}

Next, we expanded the age range to 6--13.5~Gyr and repeated the fit
(see figure~\ref{fits}d).  The width of the red giant branch is now
matched without a mismatch at the subgiant branch ({\it inset}).  The
best-fit model can be broadly characterized by a combination of two
dominant populations (figure~\ref{sfh}): 56\% of the stars ({\it red})
are metal-rich ([Fe/H]$> -0.5$) and of intermediate age (6--11 Gyr),
while 30\% of the stars ({\it blue}) are metal-poor ([Fe/H]$< -0.5$)
and old (11--13.5 Gyr).  About half of this metal-rich population
(i.e., 28\% of the total population) is 6--8~Gyr old. Note that
these models only illustrate, in a broad sense, the dominant
populations present in the M31 halo.  Other combinations of young
metal-rich and old metal-poor stars are possible.  For example, we
produced a similar fit by combining two very distinct isochrone
groups: one at 6--8 Gyr with [Fe/H]$>-1$, and one at 11.5--13.5 Gyr
with [Fe/H]$<-1$.  We are currently investigating detailed constraints
on the star formation history.

\begin{figure}
\includegraphics{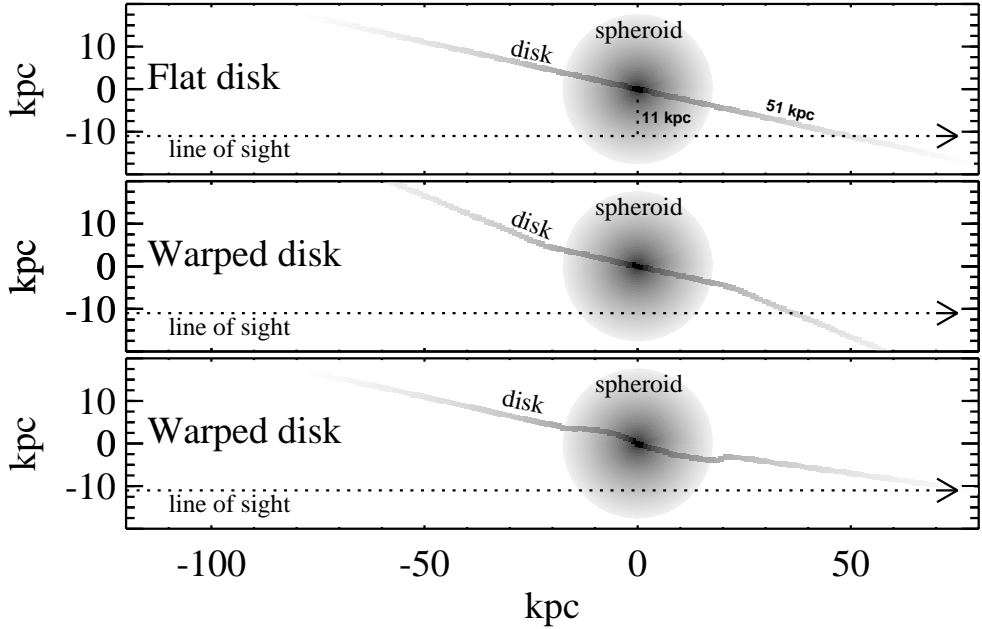}
\vspace{3.35in}
\caption{{\it Top panel:} A simple schematic showing the intersection
of our sight-line with the Andromeda disk.  The plane of the disk is
inclined $12.5^{\rm o}$ with respect to our line of sight
(\cite{dV58}), so that we are viewing the halo population at 11 kpc
from the nucleus but the disk population at 51 kpc from the nucleus.
The disk has an exponential scale length of 5.3 kpc
(\cite{WK88}). Note that the southeastern half of the disk is the side
furthest from the observer (\cite{IR85}).  {\it Middle panel:} The
isophotes for Andromeda show evidence of a warp near 22 kpc on the
major axis (\cite{WK87}).  Here, I insert a hypothetical 10$^{\rm o}$
warp at 22 kpc on the minor axis, such that more of the disk
population is moved into our sight-line.  Our sight-line has
effectively moved inward by about 2 scalelengths, increasing the disk
population by a factor of 8.  However, even with this large warp into
our sight-line, the increase in the disk population would not explain
the large population of intermediate-age stars we find in our field.
{\it Bottom panel:} A more realistic depiction of the warp in the
Andromeda disk, by interpolating the measurements of \cite[Braun
(1991)]{B91} along the minor axis.  Early models of the warp (e.g.,
\cite{H79}) imply our sight-line would not intersect the disk at all, because
of the increasing inclination at large distances from the nucleus.
Later models (e.g., \cite{B91}) indicate that the
inclination varies from 25$^{\rm o}$ close to the nucleus to
10--15$^{\rm o}$ further out, as depicted here.  The result is that
our effective position within the disk is significantly beyond the 
51 kpc shown in the top panel.}
\label{warp}
\end{figure}

\section{Disk Contamination}

Note that our halo field should be relatively free of stars moving in
the stellar disk.  Several estimates for the disk contribution at this
distance range from 1--3\%, even with a significant thick disk
component (\cite[Holland et al.\ 1996]{HFR96} and references therein).
On average, the plane of the disk is inclined from our sight-line by
$12.5^{\rm o}$ (\cite{dV58}; figure~\ref{warp}, {\it top panel}), so
we are observing halo stars 11 kpc from the nucleus but disk stars 51
kpc from the nucleus.  A warp in the outer disk, if it angled the disk
into our sight-line, would move the sight-line intersection with the
disk inward, but even a large warp could not explain the number of
young stars seen in our field.  E.g., \cite[Walterbos \& Kennicutt
(1987)]{WK87} found a possible warping of the stellar disk at
100$^\prime$ (22 kpc) on the southwest major axis.  If we assume that
the stellar disk along the southeast minor axis is warped such that
the disk beyond 22 kpc is inclined by an additional 10$^{\rm o}$ into
our sight-line, this would move our effective disk position inward by
about two scalelengths (figure~\ref{warp}, {\it middle panel}),
increasing the disk population by a factor of 8 (i.e., the disk would
contribute $\approx$10--20\% of the total population in our field).
However, we find that 56\% of the stars in our field are metal-rich
and of intermediate age, so this hypothetical increase in the disk
contamination would not account for our surprisingly young population.
In reality, studies of the neutral gas kinematics (e.g., \cite{B91};
\cite{H79}) find that the outer disk is actually viewed significantly
closer to edge-on than the inner disk (figure~\ref{warp}, {\it
bottom panel}), such that the warp in the disk would move our sight-line
further out, beyond the intersection given by a flat disk.  Finally,
the fact that the metallicity distribution in our field is very
similar to that twice as far out (\cite[Holland et al.\ 1996]{HFR96};
\cite{DHP01}) provides another strong indication that we are viewing a
halo-dominated population.

\section{Summary and Discussion}

The CMD of the M31 halo is evidently inconsistent with a population
composed solely of old (globular cluster age) stars; instead, it is
dominated by a population of metal-rich intermediate-age stars.
Although the high metallicity in the M31 halo is well-documented, the
large age spread required to simultaneously reproduce the red giant
branch, subgiant branch, and main sequence came to us as a
surprise. Earlier studies of the red giant branch were insensitive to
this age spread.  For example, \cite[Durrell et al.\ (2001)]{DHP01}
were able to explain the metallicity distribution 20 kpc from the
nucleus, with a simple chemical evolution model forming most of the
stars at very early times (see also \cite{C00}).  Although our field
is relatively small in sky coverage, it appears representative; the
metallicity in our field (\cite[Holland et al.\ 1996]{HFR96}) agrees
well with that much further out (\cite[Durrell et al.\ 2001]{DHP01}),
and there are no indications of substructure or tidal streams in the
region we surveyed (\cite{FII02}).

It seems unlikely that star formation in the halo proceeded for
$\sim$6~Gyr from gas in situ; instead, the broad age dispersion in the
halo is likely due to contamination from the disruption of satellites
or of disk material into the halo during mergers.  Indeed, our current
analysis of the data cannot rule out the possibility that M31 and a
nearly-equal-mass companion galaxy experienced a violent merger when
the Universe was half its present age.  If the 6--8 Gyr population in
the halo represents the remnants of a disrupted satellite, the
relatively high metallicity suggests that it must have been fairly
massive.  On the other hand, the stars could represent disruption of
the M31 disk, either by a major collision when M31 was $\sim 6$ Gyr
old, or by repeated encounters with smaller satellites.  The resulting
halo would be a mix of the old metal-poor stars formed earliest in
M31's halo, disk stars that formed prior to the merger(s) that were
subsequently dispersed into the halo, stars formed during the
merger(s), and the remnant populations of the disrupted satellite(s). 

\begin{acknowledgments}
This work was done in collaboration with H.C.\ Ferguson, E.\ Smith
(STScI), R.A.\ Kimble, A.V.\ Sweigart (NASA/GSFC), R.M.\ Rich, D.\
Reitzel (UCLA), A.\ Renzini (ESO), and D.A.\ VandenBerg (U.\ of
Victoria).  I am grateful to J. Harris (STScI) and P. Stetson (DAO)
for respectively providing the StarFish and DAOPHOT-II codes and
assistance in their use.  Thanks to the members of the scheduling and
operations teams at STScI (especially P.\ Royle, D.\ Taylor, and D.\
Soderblom) for their efforts in executing a large program during a
busy HST cycle.  Support for program 9453 was provided by NASA through
a grant from the Space Telescope Science Institute, which is operated
by the Association of Universities for Research in Astronomy, Inc.,
under NASA contract NAS 5-26555.  The Digitized Sky Survey was
produced at the Space Telescope Science Institute under
U.S. Government grant NAG W-2166. The images of these surveys are
based on photographic data obtained using the Oschin Schmidt Telescope
on Palomar Mountain and the UK Schmidt Telescope. The plates were
processed into the present compressed digital form with the permission
of these institutions.  This research has made use of the SIMBAD
database, operated at CDS, Strasbourg, France.
\end{acknowledgments}

\end{document}